\documentclass[prd, preprintnumbers, amsmath, amssymb, nofootinbib, aps, superscriptaddress,]{revtex4-2}

\usepackage{bm}
\usepackage{amsfonts}
\usepackage{mathrsfs}
\usepackage{graphicx}
\usepackage{amsmath}
\usepackage{float}
\usepackage{braket}

\begin{document}

\newcommand{\sgn}{\operatorname{sgn}}
\newcommand{\hhat}[1]{\hat {\hat{#1}}}
\newcommand{\pslash}[1]{#1\llap{\sl/}}
\newcommand{\kslash}[1]{\rlap{\sl/}#1}
\newcommand{\lab}[1]{}
\newcommand{\iref}[2]{}
\newcommand{\sto}[1]{\begin{center} \textit{#1} \end{center}}
\newcommand{\rf}[1]{{\color{blue}[\textit{#1}]}}
\newcommand{\eml}[1]{#1}

\newcommand{\er}[1]{Eq.\eqref{#1}}
\newcommand{\df}[1]{\textbf{#1}}
\newcommand{\mdf}[1]{\pmb{#1}}
\newcommand{\ft}[1]{\footnote{#1}}
\newcommand{\n}[1]{$#1$}
\newcommand{\fals}[1]{$^\times$ #1}
\newcommand{\new}{{\color{red}$^{NEW}$ }}
\newcommand{\ci}[1]{}
\newcommand{\de}[1]{{\color{green}\underline{#1}}}
\newcommand{\ke}{\rangle}
\newcommand{\br}{\langle}
\newcommand{\lb}{\left(}
\newcommand{\rb}{\right)}
\newcommand{\lbk}{\left[}
\newcommand{\rbk}{\right]}
\newcommand{\blb}{\Big(}
\newcommand{\brb}{\Big)}
\newcommand{\nn}{\nonumber \\}
\newcommand{\p}{\partial}
\newcommand{\pd}[1]{\frac {\partial} {\partial #1}}
\newcommand{\cd}{\nabla}
\newcommand{\cc}{$>$}
\newcommand{\bqa}{\begin{eqnarray}}
\newcommand{\eqa}{\end{eqnarray}}
\newcommand{\bqe}{\begin{equation}}
\newcommand{\eqe}{\end{equation}}
\newcommand{\bay}[1]{\left(\begin{array}{#1}}
\newcommand{\eay}{\end{array}\right)}
\newcommand{\eg}{\textit{e.g.} }
\newcommand{\ie}{\textit{i.e.}, }
\newcommand{\iv}[1]{{#1}^{-1}}
\newcommand{\st}[1]{|#1\ke}
\newcommand{\at}[1]{{\Big|}_{#1}}
\newcommand{\zt}[1]{\texttt{#1}}
\newcommand{\non}{\nonumber}
\newcommand{\m}{\mu}

\def\xa{{m}}
\def\xA{{m}}
\def\xb{{\beta}}
\def\xB{{\Beta}}
\def\xd{{\delta}}
\def\xD{{\Delta}}
\def\xe{{\epsilon}}
\def\xE{{\Epsilon}}
\def\xve{{\varepsilon}}
\def\xg{{\gamma}}
\def\xG{{\Gamma}}
\def\xk{{\kappa}}
\def\xK{{\Kappa}}
\def\xl{{\lambda}}
\def\xL{{\Lambda}}
\def\xo{{\omega}}
\def\xO{{\Omega}}
\def\xvp{{\varphi}}
\def\xs{{\sigma}}
\def\xS{{\Sigma}}
\def\xt{{\theta}}
\def\xvt{{\vartheta}}
\def\xT{{\Theta}}
\def \Tr {{\rm Tr}}
\def\CA{{\cal A}}
\def\CC{{\cal C}}
\def\CD{{\cal D}}
\def\CE{{\cal E}}
\def\CF{{\cal F}}
\def\CH{{\cal H}}
\def\CJ{{\cal J}}
\def\CK{{\cal K}}
\def\CL{{\cal L}}
\def\CM{{\cal M}}
\def\CN{{\cal N}}
\def\CO{{\cal O}}
\def\CP{{\cal P}}
\def\CQ{{\cal Q}}
\def\CR{{\cal R}}
\def\CS{{\cal S}}
\def\CT{{\cal T}}
\def\CV{{\cal V}}
\def\CW{{\cal W}}
\def\CY{{\cal Y}}
\def\BC{\mathbb{C}}
\def\BR{\mathbb{R}}
\def\BZ{\mathbb{Z}}
\def\sA{\mathscr{A}}
\def\sB{\mathscr{B}}
\def\sF{\mathscr{F}}
\def\sG{\mathscr{G}}
\def\sH{\mathscr{H}}
\def\sJ{\mathscr{J}}
\def\sL{\mathscr{L}}
\def\sM{\mathscr{M}}
\def\sN{\mathscr{N}}
\def\sO{\mathscr{O}}
\def\sP{\mathscr{P}}
\def\sR{\mathscr{R}}
\def\sQ{\mathscr{Q}}
\def\sS{\mathscr{S}}
\def\sX{\mathscr{X}}

\def\slz{SL(2,Z)}
\def\slr{$SL(2,R)\times SL(2,R)$ }
\def\ads{${AdS}_5\times {S}^5$ }
\def\adst{${AdS}_3$ }
\def\sun{SU(N)}
\def\ad#1#2{{\frac \delta {\delta\sigma^{#1}} (#2)}}
\def\bqf{\bar Q_{\bar f}}
\def\nf{N_f}
\def\sunf{SU(N_f)}

\def\dcirc{{^\circ_\circ}}

\author{Morgan H. Lynch}
\email{morganlynch1984@gmail.com}
\affiliation{Department of Electrical Engineering,
Technion: Israel Institute of Technology, Haifa 32000, Israel}
\affiliation{Center for Theoretical Physics,
Seoul National University, \\ Seoul 08826, Korea}

\title{Gravitational radiation with kinetic recoil}
\date{\today}

\begin{abstract}
In this manuscript, we examine the gravitational radiation emitted by binary systems using an Unruh-DeWitt detector coupled to gravitons. Recoil is incorporated into the system via a kinetic energy term in the energy gap of the detector. We find a splitting of the gravitational wave frequency due to the recoil. Implications for the recoil velocity and force are discussed. 
\end{abstract}


\maketitle
\section{Introduction}

The emission, and detection, of gravitational waves due to binary inspiral has ushered in the era of graviational wave astronomy \cite{abbott}. When black holes, or other compact objects, are in a binary orbit, the emission of gravitational radiation causes the decay of the orbit and eventual merger of the two objects. When this emission is asymmetric, linear momentum is radiated away by the system which then imparts a kick or recoil velocity onto the binary system or final state compact object \cite{peres, bekenstein, fitchett}. The nature of this recoil or radiation reaction poses an interesting problem in the dynamics of gravitational wave emission due to the fact that the recoil velocity may also be large enough to provide the necessary escape velocity to eject the remnant from its host galaxy, $v \sim 500 \;$ km/s \cite{varma, holz}; especially in the case of precessing binaries. Given that the properties of a gravitational wave signal encodes the properties of the binary system \cite{jolien}, e.g. gravitational wave frequency yields the orbital frequency, chirps determine the luminosity distance, etc., one can ask if there is also a signature of recoil in the measured signal of gravitational waves. The catalog of gravitational wave sources demonstrates that the mass radiated during the inspiral is typically $1 - 10 \%$ of the total mass of the system \cite{abbott1}. Such large fractions of total mass of the system radiated apriori implies the presence of radiation reaction or recoil. Thus it stands to reason that from a purely observational point of view, the parameter space of gravitational wave signals would imply that recoil would play a role in they dynamics which is on par with other characteristics of binary systems such as spin and eccentricity.  

The problem of radiation reaction in gravitational wave emission is, modulo differences due to the polarization of vectorial and tensorial modes, equivalent to that of photon emission \cite{peres}. Broadly speaking, when the energy of emitted radiation becomes comparable to the rest mass of the emitting particle, one expects a measurable presence of recoil to occur. As such, the recent experimental observation of radiation reaction at CERN-NA63 \cite{wistisen, lynch} offers us a striking window into how to incorporate radiation reaction into the problem of binary inspiral. Of particular use in gaining an insight into recoil is the Unruh-DeWitt detector \cite{lynch, lynch1}. There, the incorporation of radiation reaction is accomplished by simply including the recoil kinetic energy of photon emission into the energy gap of the detector. Moreover, the utility of the Unruh-DeWitt detector as a model for a classical radiating source has also been firmly established \cite{gapless}. This also applies to composite objects \cite{muller, matsas1, matsas3, lynch2, lynch3} such as hadrons, atoms, and, as we will demonstrate, gravitationally coupled binary systems. Then, by adapting the Unruh-DeWitt detector formalism to incorporate the emission of gravitons, we can also examine the recoil produced in these gravitational setting. This not only gives us a foothold into analyzing binary inspiral with recoil, but also extends the already robust arena of quantum field theory in curved spacetime \cite{parker, davies, aspects, pt} to include graviton emission from Unruh-DeWitt detectors.

The power radiated by binary inspiral serves as a benchmark in the observation and analysis of gravitational wave detection. In particular, the power emitted from point masses in a binary orbit describes the measured signals, in the classical regime, quite well. Known as the Peters-Mathews equation \cite{pm}, we will reproduce this result using an Unruh-DeWitt detector and, with the added degree of freedom of the energy gap, incorporate recoil into the analysis. This will note only enable us to look for signatures of recoil in the measured gravitational wave signals but also give us insight into the recoil velocities and forces produced by the radiation reaction.  We will start by computing the graviton response function and then apply the formalism to the standard binary system of gravitational wave emission. Here and throughout, unless otherwise stated, we use natural units $\hbar = c = G = 1$.

\section{The Graviton Response Function}

To begin our analysis we must first define the graviton response function, i.e. the transition rate of an Unruh-DeWitt detector coupled to gravitons. The Unruh-DeWitt detector \cite{unruh1, dewitt} will be coupled to the energy momentum tensor of our emitter and will be used to incorporate any local change in energy, e.g. a recoil kinetic energy, during the emission process. In this regard we turn to our interaction action for a graviton \cite{weinberg, poddar}, $\hat{h}^{\m \nu}(x)$,  coupled to an energy momentum tensor, $\hat{T}_{\m \nu}(x)$. Thus we have,	

\bqe
\hat{S}_{I} = \frac{1}{2} \kappa \int d^{4}x  \hat{T}_{\m \nu}(x)\hat{h}^{\m \nu}(x).
\eqe

Here, our gravitational coupling is defined by $\kappa = \sqrt{32 \pi G}$ and $d^{4}x = d^{3}xdt$. We will now use this action in order to examine an energy transition in our Unruh-DeWitt detector, from $E_{i}$ to $E_{f}$, accompanied by the simultaneous emission of a graviton with momentum $\mathbf{k}$. We will then formulate the following amplitude;

\bqe
\mathcal{A} = i\bra{\mathbf{k}}\otimes \bra{E_{f}}\hat{S}_{I}\ket{E_{i}}\otimes \ket{0}.
\eqe

The differential emission probability per unit final state graviton momenta is given by, $\frac{d\mathcal{P}}{d^{3}k} = \vert \mathcal{A} \vert^{2} = \mathcal{A}(x)\mathcal{A}^{\ast}(x')$. Evaluation yields

\bqa
\frac{d\mathcal{P}}{d^{3}k} &=& \vert  \bra{\mathbf{k}}\otimes \bra{E_{f}}\frac{1}{2} \kappa\int d^{4}x \hat{T}_{\m \nu}(x)\hat{h}^{\m \nu}(x)\ket{E_{i}}\otimes \ket{0} \vert^{2} \non \\
&=& \frac{\kappa^{2}}{4}\int d^{4}x\int d^{4}x'\vert   \bra{E_{f}} \hat{T}_{\m \nu}(x)\ket{E_{i}} \vert^{2} \vert \bra{\mathbf{k}} \hat{h}^{\m \nu}(x)\ket{0} \vert^{2}.
\eqa

Note, these matrix elements are functions of both $x$ and $x'$, e.g. $ \vert \bra{E_{f}} \hat{T}_{\m \nu}(x)\ket{E_{i}} \vert^{2} =  \bra{E_{f}} \hat{T}_{\m \nu}(x)\ket{E_{i}}  \bra{E_{i}} \hat{T}^{\ast}_{\m \nu}(x')\ket{E_{f}}$. As such, the above probability factorizes into an energy momentum tensor matrix element contracted with the graviton matrix element. We will evaluate our energy momentum component first. The energy momentum tensor itself is, in principle, comprised of all potential sources of gravitation in the binary system. We will restrict our analysis to only the mass of the system, as a first approximation, and ignore additional sources such as electromagnetic fields. To this end, we shall take our energy momentum tensor to be that of a point particle, see e.g. \cite{weinberg} page 44, coupled to an Unruh-DeWitt detector. Hence,

\bqe
\hat{T}_{\m \nu}(x) = \m v_{\m}v_{\nu}\hat{m}(t)\delta^{3}(\vec{x}-\vec{x}_{tr}(t)). 
\eqe

Here, we have defined our rest mass, or reduced mass in the case of binary inspiral, of our system, $\m$. Moreover, we evolve the detector and trajectory via the coordinate time since we will be assuming a non-relativistic velocity of our binary system, i.e. $\gamma = 1$.  The energy momentum tensor is defined by the coordinate velocity of our emitter, $v_{\m}$. The monopole moment operator $\hat{m}(t)$ is Heisenberg evolved via $\hat{m}(t) = e^{i\hat{H}t}\hat{m}(0)e^{-i\hat{H} t}$ with $\hat{m}(0)$ defined as $\hat{m}(0)\ket{E_{i}} = \ket{E_{f}}$ with $E_{i}$ and $E_{f}$ the initial energy and final energy of our energy transition which accompanies the emission along the trajectory, $\vec{x}_{tr}(t)$. The matrix element for our energy momentum tensor then yields,

\bqa
\vert   \bra{E_{f}} \hat{T}_{\m \nu}(x)\ket{E_{i}} \vert^{2} &=& \vert   \bra{E_{f}} \gamma v_{\m} v_{\nu}(x)e^{i\hat{H}t}\hat{m}(0)e^{-i\hat{H} t}\delta^{3}(\vec{x}-\vec{x}(t))\ket{E_{i}} \vert^{2} \non \\
&=& \m^{2} V_{\m \nu\sigma \rho }[x',x]\delta^{3}(\vec{x}-\vec{x}_{tr}(t))\delta^{3}(\vec{x}'-\vec{x}'_{tr}(t')) e^{-i\Delta E(t'-t)} 
\label{current}
\eqa

Here we have defined the energy gap as $\Delta E = E_{f} -E_{i}$ and normalized our detector states via $\vert \bra{E_{f}} \hat{m}(0)\ket{E_{i}} \vert^{2} = 1$. We have also defined a ``velocity tensor" via $V_{\m \nu \sigma \rho }[x',x] = v_{\nu}(x)v_{\m}(x)v_{\sigma}(x')v_{\rho}(x')  $. Next, we shall evaluate the graviton matrix element. For this we will also use the integral over the final state momenta in order to so we may obtain the total emission probability. Hence,

\bqa
\int d^{3}k  \vert \bra{\mathbf{k}} \hat{h}^{\m \nu}(x)\ket{0} \vert^{2} &=& \int d^{3}k \bra{0} \hat{h}^{\dagger \sigma \rho}(x')\ket{\mathbf{k}}\bra{\mathbf{k}} \hat{h}^{\m \nu}(x)\ket{0} \non \\
&=& \bra{0} \hat{h}^{\dagger \sigma \rho}(x')\hat{h}^{\m \nu}(x)\ket{0} \non \\
&=& G^{\mu \nu \sigma \rho}[x',x].
\eqa

Note we have utilized the completeness relation, $\int dk \ket{k}\bra{k} = 1$, so we may formulate the graviton Wightman function, $G^{\mu \nu \sigma \rho}[x',x]$. The tensor indices encode the polarization of the graviton. Using our graviton two point function and the energy momentum matrix element we can formulate the transition probability. Hence,

\bqa
 \mathcal{P} &=& \frac{\kappa^{2}}{4}\int d^{3}k\int d^{4}x\int d^{4}x'\vert   \bra{E_{f}} \hat{T}_{\m \nu}(x)\ket{E_{i}} \vert^{2} \vert \bra{\mathbf{k}} \hat{h}^{\m \nu}(x)\ket{0} \vert^{2} \non\\
 &=& \frac{\kappa^{2}\m^{2}}{4} \int dt dt' e^{-i\Delta E (\tau' - 
\tau)} V_{\m \nu \sigma \rho}[t',t]G^{\m \nu \sigma \rho}[t',t] \non \\
&=& \frac{\kappa^{2}\m^{2}}{4} \int d\xi d \eta e^{-i\Delta E \xi}  V_{\m \nu \sigma \rho}[\xi,\eta]G^{\m \nu \sigma \rho}[\xi,\eta]
\eqa

Here we have transformed our integration to the difference and average time variables; $\xi = t' - t$ and $\eta = (t' + t)/2$ respectively. Finally, by formulating the transition probability per unit time, we obtain our graviton response function, $\Gamma = \frac{d\mathcal{P}}{d\eta}$. Hence,

\bqe
 \Gamma =\frac{\kappa^{2}\m^{2}}{4} \int d\xi  e^{-i\Delta E \xi}  V_{\m \nu \sigma \rho}[\xi,\eta]G^{\m \nu \sigma \rho}[\xi,\eta].
\eqe

Due to the fact that we must contract our 4-velocities with the polarization tensors of our graviton field, let us now explicitly examine our graviton two-point function. To this end, we will use the plane wave mode decomposition for the our graviton field in the transverse traceless gauge \cite{weinberg, ford},

\bqe
\hat{h}^{\m \nu}(x) = \int \frac{d^{3}k}{(2 \pi)^{3/2}} \frac{\sum_{i}\epsilon_{i}^{\m \nu}}{\sqrt{2\omega}} \lbk \hat{a}_{k}e^{i(\mathbf{k}\cdot \mathbf{x} - \omega t) } + \hat{a}^{\dagger}_{k}e^{-i(\mathbf{k}\cdot \mathbf{x} - \omega t)}  \rbk .
\eqe

The vacuum to vacuum Wightman function will then reduce to an integral over the momentum. Hence,

\bqa
\bra{0} \hat{h}^{\dagger \sigma \rho}(x')\hat{h}^{\m \nu}(x)\ket{0} &=& \bra{0} \int \frac{d^{3}k'}{(2 \pi)^{3/2}} \frac{\sum_{\lambda'}\epsilon_{\lambda'}^{'\dagger \sigma \rho}}{\sqrt{2\omega'}}\lbk \hat{a}_{k'}e^{i(\mathbf{k'}\cdot \mathbf{x'} - \omega' t') } + \hat{a}^{\dagger}_{k'}e^{-i(\mathbf{k'}\cdot \mathbf{x'} - \omega' t')}  \rbk \non \\
&\;\;\;\;\; \times& \int \frac{d^{3}k}{(2 \pi)^{3/2}} \frac{\sum_{\lambda}\epsilon_{\lambda}^{\m \nu}}{\sqrt{2\omega}} \lbk \hat{a}_{k}e^{i(\mathbf{k}\cdot \mathbf{x} - \omega t) } + \hat{a}^{\dagger}_{k}e^{-i(\mathbf{k}\cdot \mathbf{x} - \omega t)}  \rbk \ket{0} \non \\
&=& \frac{1}{(2 \pi)^{3}}\frac{1}{2} \int \frac{d^{3}k}{\omega}\sum_{\lambda \lambda'}\epsilon_{\lambda}^{ \m \nu}\epsilon_{\lambda'}^{'\dagger \sigma \rho} e^{i(\mathbf{k}\cdot \Delta \mathbf{x} - \omega (t'-t))}.
\eqa

Here we see that the graviton two point function is formally the same as a scalar field but with polarization tensors lending their indices. It is this two point function that we will evaluate along our trajectory, i.e. $\Delta \mathbf{x} \rightarrow \Delta \mathbf{x}_{tr}$. Combining all the pieces, our response function then takes the following form,  

\bqe
\Gamma = \frac{\kappa^{2}\m^{2}}{8}  \frac{1}{(2 \pi)^{3}}\int d\xi \int \frac{d^{3}k}{\omega} V e^{-i(\Delta E \xi - \mathbf{k}\cdot \Delta \mathbf{x}_{tr} + \omega \Delta t)}.
\eqe

Here we defined the velocity-polarization product $V = \sum_{\lambda \lambda'}\epsilon_{\lambda}^{ \m \nu}\epsilon_{\lambda'}^{\dagger \sigma \rho} V_{\m \nu \sigma \rho}[\xi,\eta]$ for brevity. Let us first evaluate the sum of our polarization tensors. Recalling the polarizations are real valued and will only have spatial components, we then have $\sum_{\lambda \lambda'}\epsilon_{\lambda}^{ \m \nu}\epsilon_{\lambda'}^{\dagger \sigma \rho} = \sum_{\lambda \lambda'}\epsilon^{ i j}\epsilon^{kl} $. Then we can consider the following graviton polarization identity \cite{ford},

\bqe
\sum_{\lambda \lambda'}\epsilon_{ i j}\epsilon_{k\ell} = \delta_{ik}\delta_{j\ell} +\delta_{i\ell}\delta_{jk} -\delta_{ij}\delta_{k\ell} + \hat{k}_{i}\hat{k}_{j}\hat{k}_{k}\hat{k}_{\ell} + \hat{k}_{i}\hat{k}_{j}\delta_{k\ell} + \hat{k}_{k}\hat{k}_{\ell}\delta_{ij} - \hat{k}_{i}\hat{k}_{\ell}\delta_{jk} - \hat{k}_{i}\hat{k}_{k}\delta_{j\ell} - \hat{k}_{j}\hat{k}_{\ell}\delta_{ik}- \hat{k}_{j}\hat{k}_{k}\delta_{i \ell}.
\eqe

Here we have defined the unit graviton momentum vector, $\hat{k} = (\cos{(\phi)}\sin{(\theta)}, \sin{(\phi)}\sin{(\theta)}, \cos{(\theta)})$, using the standard spherical coordinate chart. Then, by contracting the above polarization identity with our 4-velocity tensor we have,

\bqa
v_{\nu}(x)v_{\m}(x)v_{\sigma}(x')v_{\rho}(x')\sum_{\lambda \lambda'}\epsilon_{ i j}\epsilon_{k\ell} &=& 2(v\cdot v')^{2}-v^{2}v^{'2}+(v\cdot\hat{k})^{2}(v'\cdot\hat{k})^{2} \non \\
&+&(v\cdot\hat{k})^{2}v^{'2}+(v'\cdot\hat{k})^{2}v^{2}-4(v\cdot \hat{k})(v'\cdot \hat{k})(v\cdot v').
\eqa

Here we have all dot products, $v\cdot v'$ and $v^{2}$, being strictly restricted to the spatial components of the 4 velocities. The above expression encodes the dynamics of our emitter and thus depends explicitly on the trajectory. For each scenario, the above velocity-polarization contraction needs to be evaluated and then utilized in the response function. Let us now apply the above formalism to the case of binary inspiral.

\section{Binary Inspiral}

To begin our analysis of graviton emission by two orbiting compact objects, let us first make the assumption that our dynamics will be governed by \cite{poddar} the reduced mass of the system, $\m$, orbiting at radius, $a$, with frequency $\Omega$. The reduced mass of our two orbiting compact objects, $m_{1}$ and $m_{2}$ is given by $\m = \frac{m_{1}m_{2}}{m_{1}+m_{2}}$ and their orbital parameters $a$ and $\Omega$ are related by Keplers law, $a^{3}\Omega^2 = m_{1}+m_{2}$. As such, our four-velocities, for circular rotation in the $x-y$ plane will be given $v^{\m} = (1,-a\Omega \sin{(\Omega t)},a\Omega \cos{(\Omega t)},0)$. We will then have the following velocity-polarization contraction components,

\bqa
v\cdot v' &=& (a\Omega)^2(\sin{(\Omega t)}\sin{(\Omega t')}+\cos{(\Omega t)}\cos{(\Omega t')})\non \\
v^{2}  &=& (a \Omega)^2 \non \\
v^{'2} &=& (a \Omega)^2\non \\
v\cdot \hat{k} &=& -a \Omega \sin{(\Omega t)}\cos{(\phi)}\sin{(\theta)} + a \Omega \cos{(\Omega t)}\sin{(\phi)}\sin{(\theta)} \non \\
v'\cdot \hat{k} &=& -a \Omega \sin{(\Omega t')}\cos{(\phi)}\sin{(\theta)} + a \Omega \cos{(\Omega t')}\sin{(\phi)}\sin{(\theta)}
\eqa

For the following algebra, we will use the shorthand notation, $S = \sin{(\Omega t)} $, $S' = \sin{(\Omega t')}$, $C = \cos{(\Omega t)} $, and $C'=\cos{(\Omega t')} $. We will then have the following 4 velocity-polarization tensor contraction,  $V = \sum_{\lambda \lambda'}\epsilon_{\lambda}^{ \m \nu}\epsilon_{\lambda'}^{\dagger \sigma \rho} V_{\m \nu \sigma \rho}[x',x]$,

\bqa
V &=&  2(a \Omega)^4(SS'+CC')^{2} -(a \Omega)^4 \non \\
&+& (a \Omega)^4 \lbk  C\sin{(\phi)}\sin{(\theta)} -S\cos{(\phi)}\sin{(\theta)}\rbk^{2} \lbk C'\sin{(\phi)}\sin{(\theta)}  -S'\cos{(\phi)}\sin{(\theta)} \rbk^{2} \non \\
&+& (a \Omega)^4 \lbk  C\sin{(\phi)}\sin{(\theta)} -S\cos{(\phi)}\sin{(\theta)}\rbk^{2}  \non \\
&+& (a \Omega)^4  \lbk C'\sin{(\phi)}\sin{(\theta)}  -S'\cos{(\phi)}\sin{(\theta)} \rbk^{2} \non \\ &-& 4(a \Omega)^4 (SS'+CC')\lbk  C\sin{(\phi)}\sin{(\theta)} -S\cos{(\phi)}\sin{(\theta)}\rbk \lbk C'\sin{(\phi)}\sin{(\theta)}  -S'\cos{(\phi)}\sin{(\theta)} \rbk
\eqa

Here $\theta$ is the angle of graviton emission relative to the the $z$-axis. We also recall that $\Delta t = \xi$, and we will take the dipole approximation, $\Delta x_{tr} \cdot k \ll 1 $. Then writing our momentum integrals in our response function in spherical coordinates and aligning the momentum along the z-axis, we will have the following emission rate,

\bqe
\Gamma = \frac{\kappa^{2}\m^{2}}{8}  \frac{1}{(2 \pi)^{3}}\int d\xi \int d\theta d\phi d\omega \omega \sin{(\theta)} V e^{-i(\Delta E   + \omega)\xi}.
\eqe

The angular integrations over each of the polarization-velocity contraction components yields the following;

\bqa
 &&\int d\theta d\phi \sin{(\theta)} \lbk  C\sin{(\phi)}\sin{(\theta)} -S\cos{(\phi)}\sin{(\theta)}\rbk^{2} \lbk C'\sin{(\phi)}\sin{(\theta)}  -S'\cos{(\phi)}\sin{(\theta)} \rbk^{2}  = \frac{4}{15} \pi(2+ \cos{(2\Omega \xi)})  \non \\
&& \int d\theta d\phi \sin{(\theta)}  \lbk  C\sin{(\phi)}\sin{(\theta)} -S\cos{(\phi)}\sin{(\theta)}\rbk^{2}  = \frac{4}{3} \pi  \non \\
&& \int d\theta d\phi \sin{(\theta)}  \lbk  C'\sin{(\phi)}\sin{(\theta)} -S'\cos{(\phi)}\sin{(\theta)}\rbk^{2}  = \frac{4}{3} \pi   \non \\
&&\int d\theta d\phi \sin{(\theta)} \lbk  C\sin{(\phi)}\sin{(\theta)} -S\cos{(\phi)}\sin{(\theta)}\rbk \lbk C'\sin{(\phi)}\sin{(\theta)}  -S'\cos{(\phi)}\sin{(\theta)} \rbk  = \frac{4}{3} \pi \cos{(\Omega \xi)}.
 \eqa

Here we recall that our time coordinates need to be expressed in terms of the difference and average times and thus have made use of the identity, $\sin{(\Omega t)}\sin{(\Omega t')}+\cos{(\Omega t)}\cos{(\Omega t')} = \cos{(\Omega \xi)}$. As such, our angular integrations over polarization-velocity contraction yields the following form,

\bqe
\int d\theta d\phi  \sin{(\theta)} \sum_{\lambda \lambda'}\epsilon_{\lambda}^{ \m \nu}\epsilon_{\lambda'}^{\dagger \sigma \rho} V_{\m \nu \sigma \rho}[\xi,\eta]  = \frac{8 \pi}{15} (a \Omega)^4\lbk 1+ 3 \cos{(2 \Omega \xi)}   \rbk .
\eqe

Note, our response function, $\Gamma = \frac{d \mathcal{P}}{d\eta}$, is completely decoupled from the average time coordinate, $\eta$. Now, combining all pieces together yields the following graviton emission rate,

\bqa
\Gamma &=& \frac{\kappa^{2}\m^{2}}{120 \pi^2}(a \Omega)^4 \int d\xi d\omega \omega  \lbk 1+ 3 \cos{(2 \Omega \xi)}   \rbk e^{-i(\Delta E   + \omega )\xi} \non \\
&=&\frac{\kappa^{2}\m^{2}}{120 \pi^2}(a \Omega)^4 \int d\xi d\omega \omega  \lbk 1+ \frac{3}{2} \lb e^{i2 \Omega \xi} + e^{-i2 \Omega \xi} \rb  \rbk e^{-i(\Delta E   + \omega ) \xi}.
\eqa

Integration over the time, $\xi$, will yield the following three delta functions which encode the conservation of energy for the emission; $\delta_{0}(\Delta E   + \omega)$, $\delta_{-2}(-2\Omega +\Delta E   + \omega)$, and $\delta_{2}(2\Omega +\Delta E   + \omega)$. In order to formulate the total classical emission rate \cite{lynch, gapless}, we must also sum over transitions, both up and down, of the detector energy gap. Thus, we will have the following six delta functions; $\delta^{\pm}_{0}(\pm \Delta E   + \omega)$, $\delta^{\pm}_{-2}(-2\Omega \pm \Delta E   + \omega)$, and $\delta^{\pm}_{2}(2\Omega \pm \Delta E   + \omega)$. We wish to compute the total power radiated and therefore we must also weight the integration with an additional factor of frequency, $\mathcal{P} = \int \Gamma \omega  d\omega$. Finally, we recall $\kappa = \sqrt{32 \pi}$. Thus we have,

\bqa
\mathcal{P} &=& \frac{8}{15}\m^{2}(a \Omega)^4 \int  d\omega \omega^{2} \lbk \delta^{+}_{0}+ \delta^{-}_{0} +\frac{3}{2} \lb \delta^{+}_{-2} + \delta^{+}_{2} +  \delta^{-}_{-2} + \delta^{-}_{2} \rb \rbk
\eqa
 
Since we must restrict our emitted frequency to be positive, $\omega > 0$, in the limit of zero energy gap, $\Delta E \rightarrow 0$, we will then be left with the following integrals over the delta functions; $\frac{3}{2} \lb \delta^{+}_{-2} +  \delta^{-}_{-2}\rb$. These two delta functions yield the standard gravitational wave frequency of $\omega = 2\Omega \pm \Delta E$. The first term we neglected should, in principle, correspond to a gravitational wave emitted by some transient decay-like process with energy $\omega = \Delta E$, e.g. something like an echo \cite{niayesh}. As such, our total power radiated by our binary inspiral is given by,

\bqa
\mathcal{P} &=& \frac{4}{5}\m^{2}(a \Omega)^4 \int  d\omega \omega^{2}  \lbk  \delta(\omega + \Delta E-2\Omega ) +  \delta(\omega - \Delta E-2\Omega )\rbk \non \\
&=& \frac{4}{5}\m^{2}(a \Omega)^4  \lbk 8\Omega^2 + 2\Delta E^2 \rbk \non \\
&=& \frac{32}{5}\m^{2}a^4 \Omega^{6}\lbk 1  + \frac{\Delta E^2}{(2\Omega)^2}\rbk.
\eqa

What we find is precisely the Peters-Mathews \cite{pm} result with the additional contribution of some internal process, such as radiation reaction, which is gauged by the $\Delta E$ term. The limit $\Delta E \rightarrow 0$, which models classical radiating sources, reproduces the Peters-Mathews formula identically. Having successfully reproduced this standard result as a sanity check, let us now turn to the problem of including recoil.

\section{Gravitational Radiation Reaction}

In order to analyze the effect of recoil on the gravitational wave emission of our binary inspiral, we shall turn to the example of recoil in photon emission. This affords us an opportunity to carry over the lessons of recoil from cherenkov, larmor, and channeling radiation to graviton emission in a manner that is backed by experiment \cite{wistisen, lynch, lynch1}. The incorporation of recoil indeed finds a natural setting via the use of Unruh-DeWitt detectors. This is due to the fact that for radiating sources, the energy gap of the detector is defined as the difference between the initial and final state energy of the system during the radiation process. In other words, given an initial system described by, say, its mass, we have $E_{i} = m$. Then, upon the emission of a quanta of radiation with energy, $\omega$, we will have a final state energy with a recoil momentum, $E_{f} = \sqrt{(\omega)^2 +m^2}$. The difference in energy, $\Delta E = E_{f}- E_{i} \approx \frac{\omega^2}{2m}$, is the recoil kinetic energy imparted on the system by the emission. Since, for binary inspiral, we are looking at very large energy gravitational waves, we will need to consider the wave as being comprised of the coherent sum of $n$ gravitons, per period of binary revolution, all of the same frequency $\omega$. Each of these gravitons will contribute a kick, or recoil kinetic energy. As such, we will take our energy gap to be,

\bqe
\Delta E = \frac{n \omega^2}{2m_{r}}.
\eqe

Here we have defined the recoil mass, $m_{r}$, which we will take to be the final mass of the system, i.e. the remnant mass. Let us now return to the power radiated by our binary system. We will set the energy gap equal to the recoil kinetic energy and sum over both transitions up and down, i.e. $\Delta E = \pm \frac{n \omega^2}{2 m_{r}}$. These kinetic energies will then be used in the same delta functions which reproduces the Peters-Mathews result, i.e. $\delta^{+}_{-2}$ and $\delta^{-}_{-2}$. As such, our total power of emission will then be comprised of the two following frequencies,

\bqa
\delta^{\pm}_{-2}(\omega  \pm \frac{n \omega^2}{2 m_{r}}  - 2\Omega) & \Rightarrow &  \; \omega_{\pm} = \frac{m_{r}}{n}\lbk\mp 1 \pm \lbk 1\pm\frac{4n \Omega}{m_{r}} \rbk^{1/2}  \rbk. 
\eqa

Note, we have found that the presence of recoil has split the measured gravitational wave frequency from fundamental frequency $2\Omega$. To leading order, the recoil correction takes the form, $\omega_{\pm} \approx 2\Omega\lbk 1 \mp \frac{n\Omega}{m_{r}} \rbk$, see Fig. 1 below for the frequency splitting applied to a binary black hole merger comprised of masses $m_{1} = 85m_{\odot}$ and $m_{2} = 66m_{\odot}$ with $m_{\odot}$ being the standard solar mass. Note, we will use these parameters throughout the rest of the manuscript so as to model the gravitational wave observation GW190521 \cite{gw}. In order to integrate the subsequent delta functions, we will make use of the following jacobians; $\delta^{\pm}_{-2} \Rightarrow \lbk  1\pm\frac{4n \Omega}{m_{r}} \rbk^{1/2}$. Combining our pieces together, our gravitational wave power with kinetic recoil will be given by,

\bqa
\mathcal{P}_{r} &=& \frac{4}{5}\m^{2}(a \Omega)^4 \int  d\omega \omega^{2}  \lbk \delta(\omega  + \frac{n \omega^2}{2 m_{r}}  - 2\Omega)  +  \delta(\omega  - \frac{n \omega^2}{2 m_{r}}  - 2\Omega) \rbk \non \\
&=& \frac{4}{5}\m^{2}(a \Omega)^4 \int  d\omega \omega^{2}  \lbk \frac{\delta(\omega - \omega_{+})}{\lbk  1+\frac{4n \Omega}{m_{r}} \rbk^{1/2}}  +  \frac{\delta(\omega  - \omega_{-})}{\lbk  1-\frac{4n \Omega}{m_{r}} \rbk^{1/2}} \rbk \non \\
&=& \frac{4}{5}\m^{2}(a \Omega)^4  \lbk \frac{ \frac{m_{r}^2}{n^2}\lbk -1 + \lbk 1+\frac{4n \Omega}{m_{r}} \rbk^{1/2}  \rbk^2}{\lbk  1+\frac{4n \Omega}{m_{r}} \rbk^{1/2}}  +  \frac{\frac{m_{r}^2}{n^2}\lbk\ 1 - \lbk 1-\frac{4n \Omega}{m_{r}} \rbk^{1/2}  \rbk^{2}}{\lbk  1-\frac{4n \Omega}{m_{r}} \rbk^{1/2}} \rbk.
\eqa

This is our expression for the power radiated with recoil. We can simplify the above expression by defining the recoil enhancement, $f_{r}(\Omega)$. This will allow us to better understand how it relates to the Peters-Mathews equation. As such, we will have

\bqa
\mathcal{P}_{r} &=& \frac{32}{5}\m^{2}a^{4} \Omega^{6}  f_{r}(\Omega)\non \\
f_{r}(\Omega) &=& \frac{m_{r}^2}{8n^2\Omega^{2}} \lbk \frac{ \lbk -1 + \lbk 1+\frac{4n \Omega}{m_{r}} \rbk^{1/2}  \rbk^2}{\lbk  1+\frac{4n \Omega}{m_{r}} \rbk^{1/2}}  +  \frac{\lbk\ 1 - \lbk 1-\frac{4n \Omega}{m_{r}} \rbk^{1/2}  \rbk^{2}}{\lbk  1-\frac{4n \Omega}{m_{r}} \rbk^{1/2}} \rbk. 
\eqa

If we expand for small $\frac{4n \Omega}{m_{r}} $, we find the leading order recoil correction to 
$\mathcal{P}_{r} = \frac{32}{5}\m^{2}a^4 \Omega^6  \lbk 1 +  15\frac{n^2\Omega^2}{m_{r}^2} \rbk $. Note, the above power formula applies for the average power radiated each period. As such, for the number of gravitons emitted, which determines the gravitational wave amplitude, we will also take to be the number emitted in during each period \cite{graviton}, $n = \frac{\pi}{\Omega^2}\mathcal{P}$. As an estimation, we will use the Peters-Mathews result, i.e. without recoil, in this expression for $n$. Thus we will have, 

\bqe
n=\frac{32 \pi}{5}\m^{2}a^4 \Omega^{4}.
\eqe

We must also comment on the fact that the recoil correction is purely classical. Although the graviton number, with physical constants reinserted, $n =  \frac{32 \pi}{5}\frac{G}{c^5\hbar}\m^{2}a^4 \Omega^{4}$, contains a factor of $\hbar$, our above recoil term, $f_{r}(\Omega)$, is comprised of the combination, $\frac{n\hbar\Omega}{m_{r}c^{2}}$ together. This additional factor of $\hbar$ cancels the factor of $\hbar$ in the graviton number. Thus we have a purely classical expression for recoil. Finally, to better understand the effect that recoil will have on a graviational wave observation, let us turn to the time dependence of the frequency or ``chirp". Using Keplers law, $\Omega^{2}a^{3} = (m_{1}+m_{2})$, and the gravitional energy, $E =  -\frac{m_{1}m_{2}}{2a}$, we can determine the change in frequency during the inspiral. As such, we will have the following time dependence in the frequency of our gravitational wave emission,

\bqe
\frac{d \Omega}{d t} = \frac{96}{5} \frac{G^{5/3}}{c^5}\frac{m_{1}m_{2}}{(m_{1}+m_{2})^{1/3}}\Omega^{11/3}f_{r}(\Omega).
\eqe

As in the power radiated, we have the standard expression for the chirp along with the recoil correction. This can be integrated numerically to determine the frequency chirp during inspiral both with and without recoil. \\

\begin{figure}[H]
\centering  
\includegraphics[scale=.45]{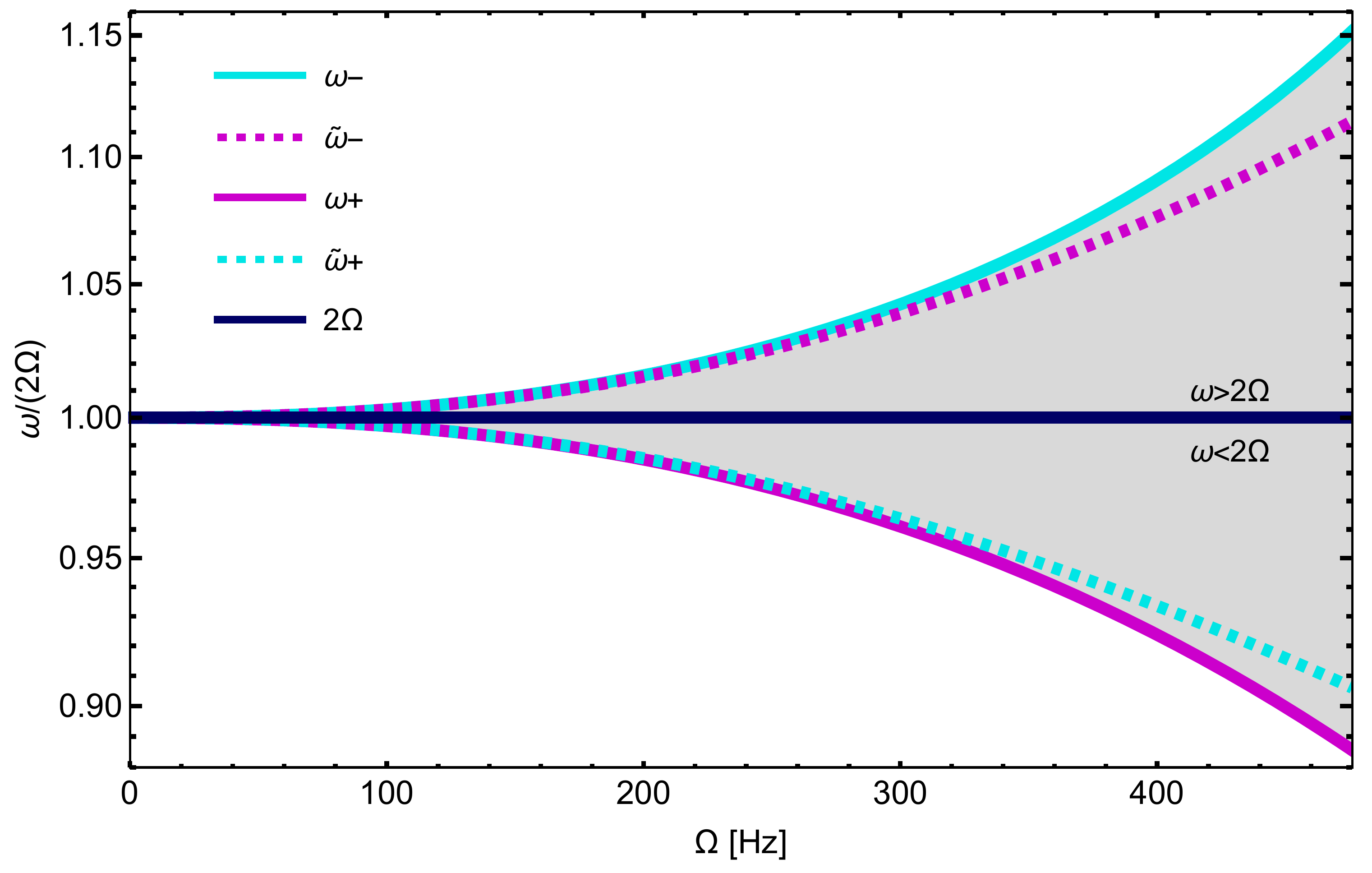}
\caption{The measured gravitational wave frequency spread, relative to the orbital frequency $\Omega$ of the binary, for $m_{1} = 85 m_{\odot}$ and $m_{2} = 66 m_{\odot}$ with the recoil mass $m_{r} = .94(m_{1} +m_{2})$. These parameters are intended to examine the observed merger GW190521 \cite{gw}. The final orbital frequency, $\sim 800$ Hz, corresponds, via Kepler's law, to a final separation determined by the Schwarzschild radii of the two initial masses. Presented are the split frequencies, Eqn. (23), for the recoil ``decay", $\omega_{-}$, and ``excitation", $\omega_{+}$, along with their approximations, $\tilde{\omega}_{-}$ and $\tilde{\omega}_{+}$}.
\end{figure}

\section{Kinematics of Recoil}

In order to analyze the kinematics of the inspiral event let us turn to the power radiated by the binary system. The Peters-Mathews result gives the power radiated, $\mathcal{P}_{PM}$, by the system without recoil present. By comparing this to the power radiated when we include the recoil correction, $\mathcal{P}_{r}$, we can estimate the amount of power that goes into accelerating the binary remnant, $\mathcal{P}_{a} = \mathcal{P}_{r} - \mathcal{P}_{pm} = \mathcal{P}_{pm}[f_{r}(\Omega)-1]$. This power, $\frac{dE}{dt}$, is then the change in kinetic energy of the remnant, i.e. $\mathcal{P}_{a} = m_{r}v\frac{dv}{dt}$. This can be integrated to yield the recoil velocity,

\bqa
v_{r} &=& \lbk  \frac{2}{m_{r}} \int \mathcal{P}_{pm} \lbk f_{r}(\Omega)-1 \rbk dt  \rbk^{1/2} \non \\
&\approx& \lbk  \frac{30}{m^3_{r}} \int \mathcal{P}_{pm} n^2(\Omega)\Omega^2 dt  \rbk^{1/2}
\eqa 

Here we have made use of the first order correction to the recoil, $ f_{r}(\Omega)-1 = 15\frac{n^2 \Omega^2}{m^2_{r}}$. We can also estimate the recoil velocity based on the spread in the measured gravitational wave frequencies measured at earth. This spread is given by the difference between the two frequencies, $\Delta \omega = \omega_{-} - \omega_{+} \approx \frac{4n \Omega^2}{m_{r}}$. From this, we have our recoil correction $ f_{r}(\Omega)-1 = \frac{15}{4}\lb \frac{\Delta \omega}{\omega_{0}} \rb^2$. Note, here we defined the fundamental frequency $\omega_{0} = 2 \Omega$. Since the presence of recoil should only manifest at the very end of the inspiral event, we can take $ \frac{\Delta \omega}{\omega_{0}} $ to be constant throughout the integration and only consider the contribution from the last few orbits at peak frequency; this criterion will then be used to define the recoil time, $t_{r} = \frac{1}{\omega_{0}}$. Then, the integral over the power will yield the total energy radiated away scaled by the ratio of the recoil time to the total time, $E_{r}\frac{t_{r}}{t_{tot}}$. As such, we will then have the following final velocity,

\bqe
v_{r} = \sqrt{\frac{15}{2}  \frac{E_{r}}{m_{r}}\frac{1}{\omega_{0}t_{tot}}}\frac{\Delta \omega}{\omega_{0}}.
\eqe

Note that, modulo binding energy, the total energy radiated and remnant mass will obey the relation, $m_{tot} = E_{rad}+m_{r}$. Also, based on the catalog of gravitational wave observations, the vast majority of energy is radiated away during the final inspiral event \cite{abbott1} and we thus take the total time of emission to only be about $t_{tot} \sim .5$ s. The utility of the above equation is that for the measured chirps of gravitational wave signals, the recoil velocity can be inferred by the ratio of the frequency broadening, $\Delta \omega$, to the frequency at maximum, $\omega$. This of course can only occur if the frequency spread due to recoil is larger than all other sources of broadening in the system, e.g. harmonics due to eccentricity, spin, and/or tidal effects. Using the same methodologies we can also examine the forces necessary to impart the final state velocity upon the remnant. From the same examination of the power imparted into the recoil, we have $\frac{dE}{dt} = v_{r}F_{r}$. As such, our force is directly proportional to the Peter-Mathews power and is given by

\bqa
F_{r} &=& \frac{ \mathcal{P}_{PM}}{v_{r}} \lbk f_{r}(\Omega) - 1   \rbk \non \\
 &=&\sqrt{\frac{15}{8 }  \frac{m_{r}}{E_{r}}\omega_{0}t_{tot}}	 \lb \frac{\Delta \omega}{\omega_{0}} \rb \mathcal{P}_{pm}.
\eqa

Then, if we take the power from Peters-Mathews at maximum frequency, $\omega_{0}$, along with the final state radius being determined by the Keplerian radius at peak frequency of the system, $a = \frac{(m_{1}+m_{2})^{1/3}}{(\omega_{0}/2)^{2/3}}$, then we have $\mathcal{P}_{PM} = \frac{1}{2^{10/3}}\frac{32}{5}\frac{(m_{1}m_{2})^2}{(m_{1}+m_{2})^{2/3}}\omega_{0}^{10/3}$. This gives us an expression for the maximum recoil force imparted on the remnant which can be inferred from the chirp signal. Thus,

\bqe
F_{r}  =\frac{1}{2^{10/3}} \frac{32}{5} \sqrt{\frac{15}{8 }  \frac{m_{r}}{E_{r}}\omega_{0}t_{tot}}	\lb \frac{\Delta \omega}{\omega_{0}} \rb \frac{ (m_{1}m_{2})^2}{ (m_{1}+m_{2})^{2/3}}\omega_{0}^{10/3}.
\eqe
 
As an example, if we examine the gravitational wave signal from event GW190521 \cite{gw}, we see the final state frequency is about $\omega_{0} \sim 70 \; s^{-1}$. Then, using Eqn. (23), we find our frequency spread to be $\Delta \omega \sim .036 \; s^{-1}$. Using $m_{r} = .94(85m_{\odot}+66m_{\odot})$ and $E_{r} = .06(85m_{\odot}+66m_{\odot})$, we then find the remnant velocity and force given by $v_{r} = 18.03$ km/s and $F_{r} = 4.04 \times10^{36}$ N or $F_{r} = 3.3\times 10^{-8}\; F_{p}$ respectively. As such, we find a recoil velocity which, although is rather large, most likely is not strong enough to eject the remnant from the host galaxy. Interestingly enough, the force imparted on the remnant to yield such a velocity is on the order of ~30 nano-Planck force, see Figures 2 and 3 below for plots of the velocities and forces for the same mass parameters as a function of binary orbital frequency. Note, these calculations were done using the approximations from Eqn.'s (29) and (31) which depend on the energy radiated, $E_{r}$, and should only be considered as an upper bound since we did not take into account things like binding energy. Using the full formulae, we find $v_{r} = 16.1$ km/s and $F_{r} = 3.7 \times 10^{-8} \; F_{p}$, which demonstrates the accuracy of the approximations employed.
 
\begin{figure}[H]
\centering  
\includegraphics[scale=.45]{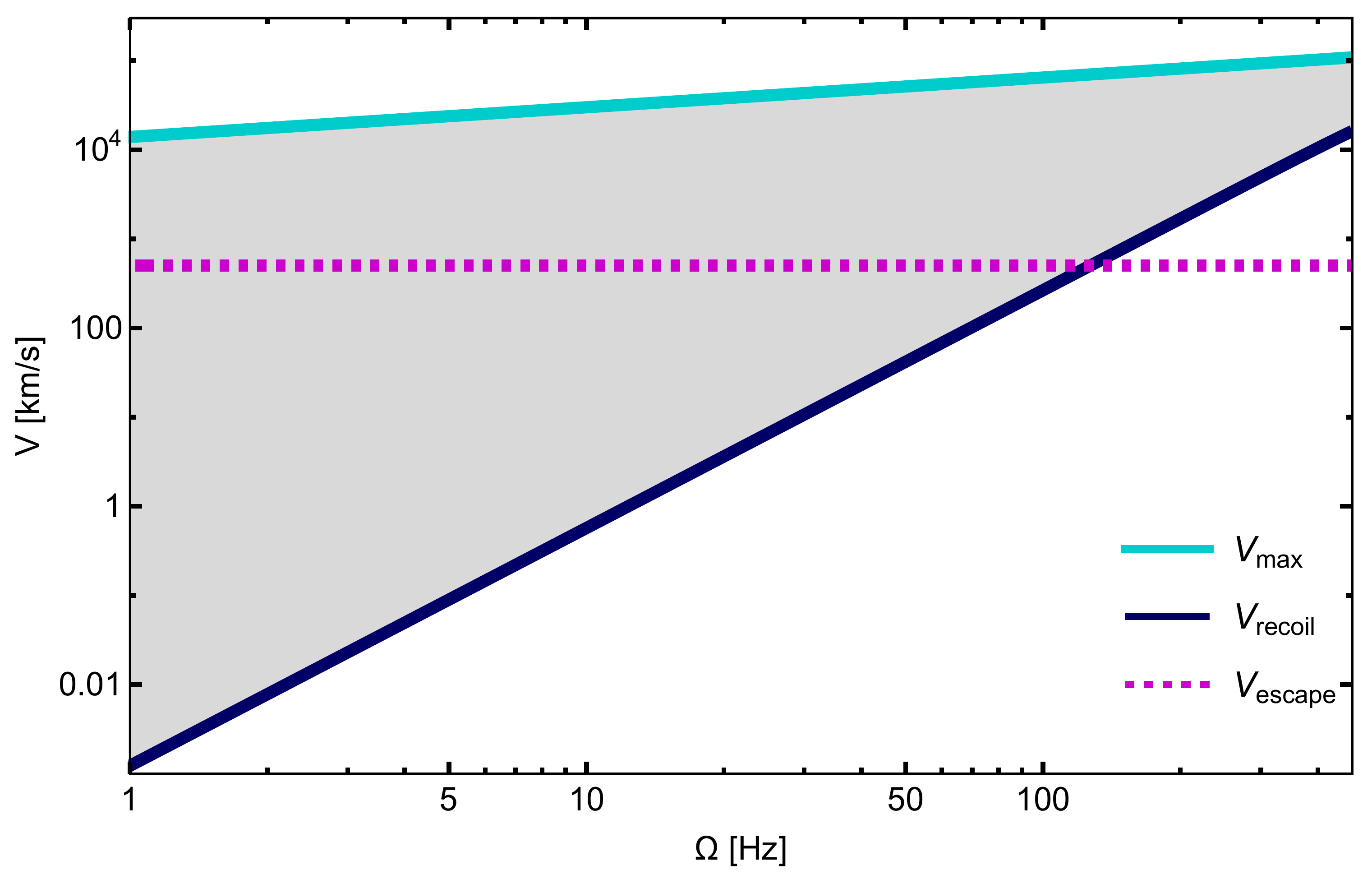}
\caption{The recoil velocity for a binary system \cite{gw} with $m_{1} = 85 m_{sun}$ and $m_{2} = 66 m_{sun}$ and recoil mass $m_{r} = .94(m_{1} +m_{2})$ as a function of the maximum binary orbital frequency. $V_{max}$ is the upper bound on the velocity for the case when all energy radiated by the system goes into accelerating the remnant. Here, we compare the computed recoil velocity to the characteristic galactic escape velocity of $V_{escape} = 500$ km/s.} 	
\label{plot3}
\end{figure}

\begin{figure}[H]
\centering  
\includegraphics[scale=.45]{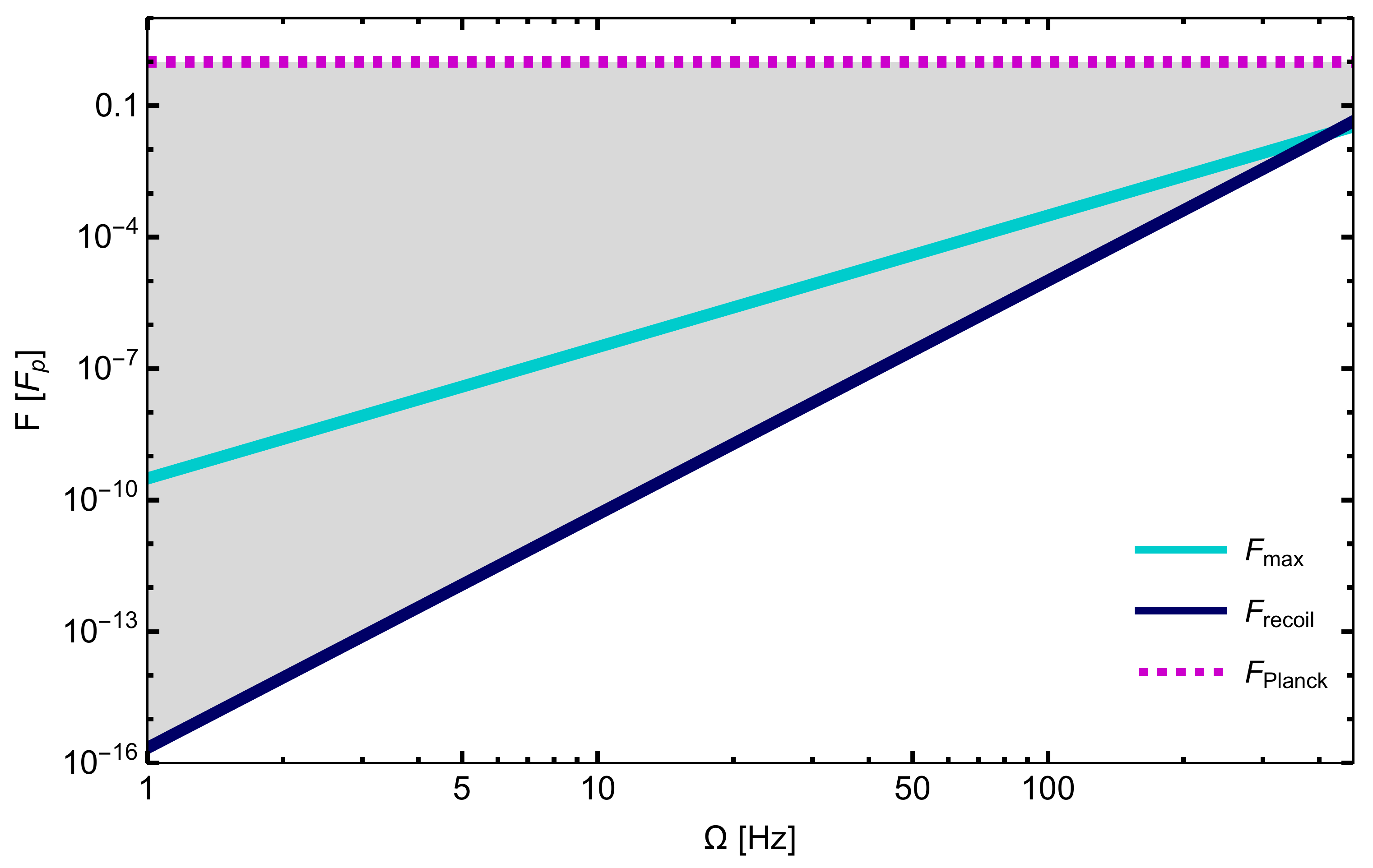}
\caption{The recoil force imparted on a binary system \cite{gw} with $m_{1} = 85 m_{sun}$ and $m_{2} = 66 m_{sun}$ and recoil mass $m_{r} = .94(m_{1} +m_{2})$ as a function of the maximum binary orbital frequency. $F_{max}$ is the upper bound on the force for the case when all energy radiated by the system goes into accelerating the remnant. We also compare the forces imparted during the recoil event to the Planck force, $F_{p} = 1.21 \times 10^{44}$ N. } 	
\label{plot3}
\end{figure}

\section{Conclusions}

In this manuscript we examined the emission of gravitational waves from binary inspiral using an Unruh-DeWitt detector coupled to gravitons. We successfully reproduced the Peters-Mathews equation as well as examined the effect of recoil on the gravitational wave frequency. We find a splitting of the fundamental frequency as a signature of radiation reaction. We also computed the final state velocity and forces present due to the recoil and find that the higher the peak frequency, the larger the remnant velocity and thus forces present. The typical forces imparted on the remnant are also on the order of $\sim$10 nano-Planck force. 

\section*{Acknowledgments}
This work has been supported by the National Research Foundation of Korea under Grants No.~2017R1A2A2A05001422 and No.~2020R1A2C2008103.

\goodbreak

\pagebreak



\begin{thebibliography}{4}

\bibitem{abbott} B. P. Abbott et al., Observation of gravitational waves from a binary black hole merger, \textit{Phys. Rev. Lett.} \textbf{116}, 061102 (2016).

\bibitem{peres} A. Peres, Classical radiation recoil, \textit{Phys. Rev.} \textbf{128}, 2471 (1962).

\bibitem{bekenstein} J. D. Bekenstein, Gravitational-radiation recoil and runaway black holes, \textit{Astrophys. J.} \textbf{183}, 657 (1973). 

\bibitem{fitchett} M. J. Fitchett, The influence of gravitational wave momentum losses on the centre of mass motion of a Newtonian binary system, \textit{Mon. Not. R. Astron. Soc.} \textbf{203}, 1049 (1982).

\bibitem{varma} V. Varma et al., Evidence of large recoil velocity from a black hole merger signal, \textit{Phys. Rev. Lett.} \textbf{128}, 191102 (2022).

\bibitem{holz} M. Favata, S. A. Hughes, and D. E. Holz, How black holes get their kicks: gravitational radiation recoil revisited, \textit{Astrophys. J.} \textbf{607}, L5 (2004).

\bibitem{jolien} J. Creighton and W. Anderson, \textit{Gravitational-Wave Physics and Astronomy}	 (Wiley-VCH, Weinheim, 2011).

\bibitem{abbott1} R. Abbott et al., GWTC-3: compact binary coalescences observed by LIGO and Virgo during the second part of the third observing run, arXiv:2111.03606 [gr-qc] (2021).

\bibitem{wistisen} T. N. Wistisen et al., Experimental evidence of quantum radiation reaction in aligned crystals, \textit{Nat. Commun.} \textbf{9}, 795 (2018).

\bibitem{lynch} M. H. Lynch et al., Experimental observation of acceleration-induced thermality, \textit{Phys. Rev. D} \textbf{104}, 025015 (2021). 

\bibitem{lynch1} M. H. Lynch et al., Accelerated-Cherenkov radiation and signatures of radiation reaction, \textit{New J. Phys.} \textbf{21}, 083038 (2019).

\bibitem{gapless} G. Cozzella et al., Uniformly accelerated classical sources as limits of Unruh-DeWitt detectors, \textit{Phys. Rev. D} \textbf{102}, 105016 (2020).

\bibitem{muller} R. Muller, Decay of accelerated particles, \textit{Phys. Rev. D} \textbf{56}, 953 (1997).

\bibitem{matsas1} G. E. A. Matsas and D. A. T. Vanzella, Decay of protons and neutrons induced by acceleration, \textit{Phys. Rev. D} \textbf{59}, 094004 (1999).

\bibitem{matsas3} D. A. T. Vanzella and G. E. A. Matsas, Decay of accelerated protons and the existence of the Fulling-Davies-Unruh effect, \textit{Phys. Rev. Lett.} \textbf{87}, 151301 (2001).

\bibitem{lynch2} M. H. Lynch, Acceleration-induced scalar field transitions of n-particle multiplicity, \textit{Phys. Rev. D} \textbf{90}, 024049 (2014).

\bibitem{lynch3} M. H. Lynch, Accelerated quantum dynamics, \textit{Phys. Rev. D} \textbf{92}, 024019 (2015).

\bibitem{parker} L. Parker, \textit{``The Creation of Particles by the Expanding Universe"}, thesis, Harvard University (1966).

\bibitem{davies} N. D. Birrell, P. C. W. Davies, \textit{Quantum Field Theory in Curved Space}	 (Cambridge University Press, Cambridge, 1982).

\bibitem{aspects} S. A. Fulling, \textit{Aspects of Quantum Field Theory in Curved Space-Time} (Cambridge University Press, Cambridge, 1989).

\bibitem{pt} L. Parker, D. Toms, \textit{Quantum Field Theory in Curved Spacetime: Quantized Fields and Gravity} (Cambridge University Press, Cambridge, 2009).

\bibitem{pm} P. C. Peters and J. Mathews, Gravitational radiation from point masses in a Keplerian Orbit, \textit{Phys. Rev.} \textbf{131}, 435 (1963).

\bibitem{unruh1} W. G. Unruh, Notes on black-hole evaporation, \textit{Phys. Rev. D} \textbf{14}, 870 (1976). 

\bibitem{dewitt} S. Hawking, W. Israel, B. S. DeWitt, \textit{General Relativity an Einstein centenary
survey}, (Cambridge University Press, Cambridge, 1979).

\bibitem{weinberg} S. Weinberg, \textit{Gravitation and Cosmology: Principles and Applications of the General Theory of Relativity} (John Wiley $\&$ Sons, New York, 1972).

\bibitem{poddar} T. K. Poddar, S. Mohanty, and S. Jana, Gravitational radiation from binary systems in massive graviton theories, \textit{J. Cosmol. Astropart. Phys.} \textbf{03}, 019 (2022).

\bibitem{ford} H. Yu and L. H. Ford, Light-cone fluctuations in flat spacetimes with nontrivial topology, \textit{Phys. Rev. D} \textbf{60}, 084023 (1999).

\bibitem{niayesh} J. Abedi, H. Dykaar, and N. Afshordi, Echoes from the abyss: Tentative evidence for Planck-scale structure at black hole horizons, \textit{Phys. Rev. D} \textbf{96}, 082004 (2017).

\bibitem{gw} R. Abbott et al., GW190521: A Binary Black Hole Merger with a Total Mass of 150 $M_{\odot}$, \textit{Phys. Rev. Lett.} \textbf{125}, 101102 (2020).

\bibitem{graviton} G. L. Murphy, Gravitons from a spinning rod, \textit{Aust. J. Phys.} \textbf{31}, 205 (1978).




 \end{thebibliography}
\end{document}